\documentclass[sn-mathphys]{sn-jnlmod}

\theoremstyle{thmstyleone}%
\theoremstyle{thmstyletwo}%

\theoremstyle{thmstylethree}%

\renewcommand{\p}{\partial}

\newcommand{\dd}{{\rm d}}

\newcommand{\tildeHor}{\widetilde{\textrm{Hor}}}

\newcounter{mnotecount}[section]

\newcommand{\mnotex}[1]
{\protect{\stepcounter{mnotecount}}$^{\mbox{\footnotesize $\bullet$\themnotecount}}$
\marginpar{
\raggedright\tiny\em
$\!\!\!\!\!\!\,\bullet$\themnotecount: #1} }

\begin{document}


\title{An anisotropic gravity theory}


\author[1]{\fnm{} \sur{A. Garc\'\i a-Parrado }}

\author[2]{\fnm{} \sur{E. Minguzzi}}

\affil[1]{\small \orgdiv{Departamento de Matem\'aticas}, \orgname{Universidad de C\'ordoba}, \orgaddress{\street{Campus de Rabanales, 14071}, \city{C\'ordoba}, \country{Spain}, 
agparrado@uco.es}}

\affil[2]{\small \orgdiv{Dipartimento di Matematica e Informatica ``U. Dini''}, \orgname{Universit\`a
	degli Studi di Firenze}, \orgaddress{\street{Via S. Marta 3,  I-50139}, \city{Firenze},  \country{Italy},
ettore.minguzzi@unifi.it}}


\abstract{
We study an action integral for Finsler gravity obtained by pulling back
an Einstein-Cartan-like Lagrangian from the tangent bundle to the base manifold. The vacuum equations are obtained imposing stationarity with respect to any section (observer) and are well posed as they are independent of the section. They imply that in vacuum the metric is actually independent of the velocity variable so the dynamics becomes coincident with that of general relativity.
}



\maketitle








%
%
%
%
%

\section{Introduction}

In this work we explore some features of a possible dynamics for Finsler gravity. The main mathematical objects of Finsler geometry, from the metric to the linear Finsler connections, live on the slit tangent bundle $E=TM\backslash 0$ of the base manifold $M$. Typically, in order to construct a Finsler action, one would integrate over the indicatrix, namely over the locus $\{(x,y): 2\mathcal{L}(x,y)=-1\}$, where $\mathcal{L}$ is the Finsler Lagrangian. This approach is followed in \cite{akbarzadeh95,chen08b,pfeifer12,shen16} particularly with reference to an action which is the integral of the Ricci scalar, see also \cite{ni19,hohmann19,javaloyes21} for further analysis of this type of action. To the best of our knowledge no investigation has been devoted to a different approach, namely to actions obtained via an integral over $M$. In general, one would need to pull back all relevant quantities to $M$ via a section $s:M\to E$ of the vector bundle $\pi: E\to M$. The problem with this approach is that the action would depend on the section. We believe that a natural solution to this problem is to impose that the dynamical equations should be equations on $E$ such that the action is stationary for every possible choice of section $s$.
The idea is that a section could represent an observer and hence the action could be observer dependent while the dynamical equations would not. It turns out that more is true. We shall find that, at least in vacuum and on-shell, the action is actually independent of the section.

Let $\{x^\alpha\}$ be local coordinates, let $e_a=e_a^\mu(x) \p/\p x^\mu$, be a basis field on $M$, let
$\{e^a\}$ be the cobasis, and let $\{x^\alpha,y^a\}$ be local coordinates induced on $E$.
A Finsler connection $\nabla$ is a Koszul connection on the pullback\footnote{We use the pullback approach which is by now standard \cite{bao00}. However, there are other approaches that give insights on the geometry of the indicatrix, see \cite{makhmali18}.} (linear) bundle $\pi^*TM\to E$. It is
well known that it is possible to construct many canonical Finsler connections (e.g.\ Cartan, Chern,
Berwald) but that all these connections induce the same non-linear connection, namely the same
splitting $TE=VE\oplus HE$ referred to as the \emph{canonical non-linear connection}. In all these cases $HE=\ker \nabla y$ where $y=y^a e_a:E\to TM$ is the
Liouville vector field. One says that all notable Finsler connections are {\em regular}. The regularity
condition is equivalent to the fact that the forms $\omega^a=\pi^* e^a$ joined with\footnote{Later we shall
introduce the covariant exterior differential $D$ of $\nabla$ so that this equation can also be written
$\bar\omega^a=Dy^a$.} $\bar \omega^a=(\nabla y)^a$,
form a basis on $T^*M$, cf.\ \cite{minguzzi21b}.

We stress that, although we make use of a linear Finsler connection, our theory should be more properly referred to as an anisotropic theory, rather than as a Finslerian theory, as we do not assume that the metric is a vertical Hessian of some Finsler Lagrangian $\mathcal{L}$ as in (pseudo-)Finsler geometry. Also the non-linear connection will not be treated as an independent variable, rather it will be deduced from the linear Finsler connection by the imposition of the regularity condition. In this way the variations of the connections will be totally unconstrained as in our  purely metric-affine theory \cite{minguzzi20b}. \footnote{A Finsler connection is {\em strongly regular} if $ \nabla y\vert_{VE}=Id_{VE}$. All notable connection share this property, but we do not impose it as this condition would constrain the variations.}

One should not expect to determine the Finsler connection univocally. Indeed, already in the Palatini formulation of the Einstein-Hilbert Lagrangian one finds that the connection is determined up to projective transformations. In order to find a unique connection, say the Levi-Civita connection, it is necessary to impose some a priori constraint such as metricity  as in the  Einstein-Cartan formulation.

In a fully metric-affine approach, in which metric and connection are a priori independent,  one is left with the conundrum of justifying why the material part of the action does not exhibit shear hypermomentum.
In a previous work \cite{minguzzi20b} we showed that one could solve this problem by enlarging the action symmetry to what we called the {\em amplified symmetry}. In short our approach showed that one can really work with equivalence classes of connections as long as the material part of the action shares the same symmetry that determines the classes.

In a Finslerian framework one is lead to reconsider these findings in light of specific features of Finsler geometry.
One might expect the non-linear (Ehresmann) connection rather than the Finsler connection to be the physical ingredient of a Finslerian gravity theory. The theory developed in \cite{pfeifer12} does indeed share this feature as it is based on the Ricci scalar that  indeed only depends on the non-linear curvature. Still there is also the chance that the true variable of a Finslerian action could be the Finsler connection, possibly constrained in some way. If that is the case one possibly natural constraint would be the metric constraint as in the Einstein-Cartan theory (with no assumption on the torsion, which in Finsler geometry leaves up much freedom).

Our work presents at once two equivalent theories. One theory, which is obtained by removing all the `tildes' in the following expressions, imposes a priori metricity on the Finsler connection, much as in the Einstein-Cartan theory. This is not a fully metric-affine theory. The other theory is obtained by implementing in the action
the following  {\em amplified symmetry} from our previous work \cite{minguzzi20b}
\[
\omega^a{}_b\to\omega^a{}_b+ A^a{}_b, \qquad A_{ab}=A_{ba}
\]
where $\omega^a{}_b$ are the coefficients for the Finsler connection and the indices are lowered with the metric $g_{ac}$.
In this theory one really works with equivalence classes of connections and the theory is fully metric-affine.
A posteriori one can then select some specific connection from the solution class, such as the metric representative.

 In this second approach we expect that the dependence on the linear connection would be mitigated and that the true variable of the theory could end up being the non-linear connection. Whether this is the case depends on the matter Lagrangian, which might have a restricted invariance. For instance, it might impose the additional condition $A_{ab} y^b=0$.
  Then, due to the equation $HE=\ker \nabla y$, each element in a \emph{restricted amplified symmetry} class would share the same non-linear connection, which shows that the action would really depend on the non-linear connection, see also \cite{minguzzi21b} for some mathematical discussion on this restricted amplified symmetry.

  We can use the restricted
 amplified symmetry to define an equivalence relation $\nabla\sim_{g}\nabla'$ in the set of regular Finsler connections.
  The notable connections of Finsler geometry  previously
mentioned all belong to
  same class of connections under such equivalence relation.\footnote{It can be observed that the notable connections are strongly regular and that they read $\omega^a_{\ b}=\omega_{Cartan}{}^a_{\ b}+\alpha L^a_{b c} \omega^c-\beta C^a_{bc} \bar \omega^c$, for suitable choices of the constants $(\alpha,\beta)$. Here $L^a_{bc}$ is the Landsberg tensor \cite{minguzzi14c}. Indeed, we have $\textrm{Cartan}=(0,0)$,  $\textrm{Chern}=(0,1)$, $\textrm{Berwald}=(1,1)$,  $\textrm{Hashiguchi}=(1,0)$. This formula and the symmetries of the Cartan and Landsberg tensors clarify that all notable connections belong to the same restricted amplified symmetry  class.}


%
%
%
%

The implementation of such an amplified symmetry means modifying the Einstein-Cartan action on $E$ to allow for non-metricity.
The way we implement the amplified symmetry is by using the following {\em metric representative Finsler connection} $\tilde \nabla$
\begin{equation} \label{byq}
\tilde \omega^a{}_b:=\omega^a{}_b+\frac{1}{2} g^{ar} D g_{rb}
\end{equation}
where $D$ is the covariant exterior differential induced by $\nabla$. Similarly, we shall denote with
$\tilde D$ the covariant exterior differential induced by $\tilde \nabla$, and in general any object
dependent on $\tilde \nabla$ will carry a tilde.
The connection $\tilde \nabla$ is indeed metric, hence $\tilde D g_{ab}=0$, and
its curvature 2-form $\tilde R$ is related to that of $\nabla$, $R$, by
\begin{equation}
 \tilde R_{ab}=R_{[ab]}-\tfrac{1}{4} D g_{ac} \wedge g^{cs} D g_{sb}.
 \label{eq:tilde-R}
\end{equation}
When $\nabla$ is one of the notable Finsler connections, $\tilde \nabla$ is the Cartan connection.\footnote{With reference to the previous footnote, a simple calculation gives $(D g)_{ab}= -2 \alpha L_{abc} \omega^c+2\beta C_{abc} \bar \omega^c$ from which the claim easily follows from Eq.\ (\ref{byq}).}



\label{sec:grav-action}
The gravitational action studied in this work is
\begin{equation} \label{bod}
S_G(g_{ab},\omega^a{}_b, e^a,s):=\int_M s^*(L_G), \qquad L_G=  \tfrac{1}{2} \eta_{abcd} \tilde R^{ab} \wedge \omega^c \wedge \omega^d,
\end{equation}
where $\omega^a=\pi^* e^a$ and where $\eta_{abcd}=\sqrt{\vert \det g(x,y)\vert} [abcd]$ is the Finslerian volume tensor.
We shall restrict ourselves to the 4-dimensional spacetime case, but the generalization to different dimensions will be obvious.

As mentioned, although we shall work with the amplified symmetry, much in analogy with \cite{minguzzi20b}, the paper also includes the case in which
this symmetry is not implemented which is recovered by removing all the `tildes'  and by assuming that $\nabla$ is metric (but a priori
not necessarily Cartan's).

Except for the pullback $s^*$, and the use of the connection $\tilde \nabla$
to define the curvature, the Lagrangian is Einstein-Cartan's. In the Einstein-Cartan theory one obtains that torsion
can be present only inside matter, and in fact, perhaps not surprisingly, we shall obtain a similar result in the Finslerian case.
However, all Finsler connections have some form of torsion unless the space is really pseudo-Riemannian,
so ultimately we shall obtain that in vacuum the spacetime is pseudo-Riemannian, i.e.\ $g$ does not depend on the fiber variables $y$.

\subsection{Energy-momentum conservation}

Independently of the previous discussion, there is another line of thought that brought us to consider a Lagrangian of the above form.
A well-known problem in Finsler geometry is that of establishing some form of conservation principle. The conservation of the stress-energy tensor was indeed one of the demands that led Einstein to the development of the general theory of relativity \cite{misner73,einstein16}.
Due to the abundance of potential field equations in Finsler gravity, a similar criteria could guide the selection of the correct ones.

We do not think, however, that in the Finslerian case one should search for a conserved stress-energy tensor but rather,
as suggested in \cite[Remark 19]{minguzzi14c}, for a conserved observer-dependent energy-momentum vector, or equivalently, 3-form (for other recent approaches and formalisms see \cite{hohmann22,javaloyes22}).

In this regard we observe that the following modified Einstein's vector-valued 3-form
\begin{equation}
\tilde \tau_d:=\frac{1}{2}\frac{\delta L_G }{\delta\omega^d}=\tfrac{1}{2} \eta_{abcd} \tilde R^{ab} \wedge \omega^c
\label{eq:tilde-tau}
\end{equation}

satisfies $\tilde \tau_d \wedge \omega^d=L_G$ and
\begin{equation} \label{bud}
\tilde D\tilde \tau_d= \tfrac{1}{2} \eta_{abcd} \tilde R^{ab} \wedge \Psi^c
\end{equation}
where, denoting with $T^c$ the horizontal torsion for $D$,
\begin{equation}
\Psi^c:=\tilde D \omega^c= T^c +  \tfrac{1}{2}
 g^{cr} Dg_{rb} \wedge \omega^b
 \label{eq:define-Psi}
\end{equation}
is the horizontal torsion of the connection $\tilde D$. Recalling that $\bar \omega^c= D y^c$ (which jointly with $\omega^c$ form a basis for $T^*E$) and similarly ${\tilde{\omega}}^c=\tilde D y^c$, we have, contracting  (\ref{bud}) with the Liouville vector field
\[
\dd (\tilde \tau_d y^d)= \tfrac{1}{2} \eta_{abcd}y^d \tilde R^{ab} \wedge \Psi^c-\tilde \tau_d\wedge{\tilde{\omega}}^d.
\]

Every form can be expressed as a linear combination of wedge products of basis forms $\{\omega^c,\bar \omega^c\}$,
so leading to various decompositions in horizontal-vertical parts. A similar statement applies if we take the basis
$\{\omega^c,{\tilde{\omega}}^c\}$. We denote with $\textrm{Hor}$  the operator that sends all the vertical forms to zero,
in other words $\textrm{Hor}$ keeps only the terms that have a totally horizontal expansion. Similarly
$\widetilde{\textrm{Hor}}$ does the same, but for the non-linear connection induced by  $\tilde \nabla$.

The nice fact is that for a notable connection $\Psi^a= C^a{}_{bc} {\tilde{\omega}}^c\wedge \omega^b$
where $C_{abc}=\tfrac{1}{2}\frac{\p}{\p y^a} g_{bc}$ is the Cartan torsion, and hence $ \tildeHor(\Psi^a)=0$.
Whenever this equation holds the previous expression in display satisfies
\[
\tildeHor(\dd (\tilde \tau_d y^d))=0.
\]

When we pullback with a section $s:M\to E$, we can make use of the identity $s^*(\tilde \omega^c)=:\tilde Ds^c$
where, with some abuse of notation, we denote with $\tilde Ds^c$ the non-linear covariant derivative of the section. This non-linear
covariant derivative is a 1-form
that vanishes at a point precisely when the section represents a free falling non-rotating non-expanding
reference frame (observer) at the point.
Such observers exist at every point  due to the fact that for every chosen point of $E$, linear coordinates can be chosen so that the
connection coefficients vanish at the point \cite{minguzzi15f}.

We conclude that at that point and for such a frame the 3-form $s^*(\tilde \tau_d y^d)$ is closed
\begin{equation} \label{bid}
\dd (s^*(\tilde\tau_d y^d))=0.
\end{equation}

Now assume dynamical equations of the form
\begin{equation} \label{vis}
\tilde \tau_d=t_d,
\end{equation}
or more weakly $y^d \tilde \tau_d=y^d t_d$, where  $t_d$ is the energy-momentum 3-form and
$y^d t_d$ is the energy-momentum of matter as viewed by an observer with velocity of direction $y^d$.

Observe that equation (\ref{bid}), now written as $\dd (s^*(t_d y^d))=0$,
 can be attained at any chosen point through a suitable choice of observer. However, in general,
 this cannot be obtained in a neighborhood of the point since a section
 $s: M \rightarrow E$ such that
 $\tilde Ds=0$ does not exist locally.
 Still the conservation can be approximately expressed in integral form in a small neighborhood of the point where the previous equation  holds true. So for a cylindrical neighborhood $C$, with basis $B_1$, $B_2$ and lateral side $S$, we shall have
\[
\int_{B_1} s^*(t_d y^d)+ \int_{B_2} s^*(t_d y^d)+\int_{S} s^*(t_d y^d)=o(\textrm{size of cylinder}) .
\]
Using the orientation coming from the time orientation of the spacetime, one of the two first terms of the previous expression changes sign, and we get an expression which states that the energy-momentum  $s^*(t_d) s^d$ entering $C$ from the basis $B_1$ is equal to that escaping $B_2$ plus that entering the lateral side $S$. The contraction with $s^d$ expresses the fact that the quantity $s^*(t_d) s^d$  is an energy-momentum 3-form (dually a vector) dependent on the chosen observer $s$ rather than a tensor. This analysis is compatible and in the same spirit of that performed in \cite[Remark 19]{minguzzi14c}, but here we do not require that the  Landsberg tensor or mean Cartan torsion vanish.


In the next section we shall see that our action varied with respect to the vierbein indeed gives an equation of the form $\tilde \tau_d=t_d$.
Here, we mention that in the pseudo-Riemannian case, with $\tilde \nabla$ the Levi-Civita connection, we indeed recover from the previous formulas the expected results for general relativity. In this case $\tilde{\tau}_d$ becomes
\begin{equation} \label{hhv}
\tilde \tau_d=\tfrac{1}{6} G_d{}^r \eta_{ruvz} \omega^u
\wedge \omega^v \wedge \omega^z
\end{equation}
where $G_{ab}=R_{ab}-\tfrac{1}{2} R g_{ab}$ is the Einstein tensor and the energy-momentum 3-form $t_a$
is obtained from a suitable matter Lagrangian ${L}_M$ by
\begin{equation}
t_a=-\frac{1}{2}\frac{\delta {L}_{M} }{\delta\omega^a}=
\frac{1}{6} T_{d }{}^{r} \eta_{r uvz} \omega^u \wedge\omega^v \wedge\omega^z.
\label{eq:matter-3-form}
\end{equation}
The dynamical equation \eqref{vis} becomes in this case
\begin{equation}
G_{ab}=T_{ab},
\end{equation}
which implies the standard relation $\tilde{\nabla}_a T^{a}{}_b=0$.
Observe that in general relativity, if $u^a$ is a 4-velocity field,
and $T_{ab}$ is the matter energy-momentum tensor defined by \eqref{eq:matter-3-form}, then
$\tilde{\nabla}_a(T^{ab} u_b)=T^{ab}\tilde{\nabla}_bu_{a}\ne 0$.
Thus, also according to the anisotropic theory, conservation of energy-momentum is expected only in free-falling non-rotating non-expanding reference frames and only locally as both features are shared by classical general relativity.

\subsection{Action variation and equations of motion}

In this section we shall investigate whether it is possible to obtain the dynamical equation (\ref{vis}) through the variation of an action. We shall be concerned in recovering just the left-hand side of  (\ref{vis})  which has gravitational origin.

Let  $T^a= D \omega^a$ and $\bar T^a=D\bar \omega^a$ be the horizontal and vertical torsions respectively, and let $\mathcal{G}_{ab}=D
g_{ab}$ be the non-metricity, all with reference to the same Finsler connection $\nabla$. It is useful to recall the next Bianchi
identities which hold for any Finsler connection (including $\tilde D$ which is metric, hence the last one
establishes the antisymmetry of $\tilde R_{ab}$)
\begin{align*}
D R^a{}_b&=0, \\
 D T^a&= R^a{}_b \wedge \omega^b, \\
 D \bar T^a&= R^a{}_b \wedge \bar \omega^b, \\
 D \mathcal{G}_{ab}&=-R_{ab}-R_{ba}.
 \end{align*}

Let us consider the gravitational Einstein-Cartan-like action (\ref{bod}).
We shall often use the following observation. By using normal coordinates at a point of $E$ it is easy to check that given a non-
vanishing $k$-vector on $TM$, and a $k$-vector on $TE$ that projects on it, it is always possible to find a section $s:M\to E$ such that
the former is sent to the latter (in normal coordinates it is sufficient to take $s^a=A^a_{b} x^b$ where $A$ is a suitable matrix).
Since any $k$-vector on $TE$ can be approximated by a $k$-vector that projects to a non-vanishing $k$-vector we have, by continuity, that
given any $k$-form $\omega$ on $E$, the equality $s^*\omega=0$ holds for every $s$ iff $\omega=0$ holds
(see \ref{subsec:appendix-sstar} in the Appendix for a detailed derivation). This property shall explain why
the pullback can be omitted from our equations, as they are supposed to hold for every section.

Another useful fact, that  we shall systematically use in our calculations,
is the commutativity of the exterior derivative with the pullback $\dd s^*\alpha=s^* \dd \alpha$, where $\alpha$ is a form on $E$.
Ultimately, it will allow us to get rid of some exact terms.

The \emph{configuration variables} of the action $S_G$ are $\omega^a$, $\omega^a{}_b$ and $g_{ab}$. This means that we can define the action variation
with respect to all of them in the usual way
\begin{equation}
 \delta S_G:= \left.\frac{d}{d t} S_G((\omega^t)^a,\; (\omega^t)^a{}_b,\; g^t_{ab})\right\lvert_{t=0},
\label{eq:delta-variation}
\end{equation}
where the subscript $t$ is a real parameter that defines the \emph{variation fields}. The previous
definition is generalized in the obvious way to define the variation of any quantity with respect to a
selected set of configuration variables.
The general variation $\delta S_G$ for fixed section $s$ can be written, using $s^*\pi^* e^a=e^a$ and the standard integration by parts,
as
\[
\delta S_G=\int_M s^*\left(\frac{\delta L_G}{\delta \omega^a}\right) \wedge \delta e^a+
s^*\left(\frac{\delta L_G}{\delta \omega^a{}_b} \wedge \delta \omega^a{}_b\right)
+  s^*\left(\frac{\delta L_G}{\delta g^{ab}}  \delta g^{ab}\right) + \int_M s^*(d \Xi).
\]
The last term can be removed using the Stokes theorem and suitable boundary conditions.
Finally, we arrive at (see Appendix B)
\begin{align}
\frac{\delta L_G}{\delta \omega^d}&=2\tilde \tau_d=\eta_{abcd} \tilde R^{ab} \wedge \omega^c, \label{one}\\
\frac{\delta L_G}{\delta \omega^a{}_r} &= - g^{rb}\eta_{abcd}   \Psi^c  \wedge \omega^d ,\label{two}\\
\frac{\delta L_G}{\delta g^{ab}}&= -\frac{\delta L_G}{\delta g_{mn}} g_{ma} g_{nb}=  \nonumber \\
&\quad -\tfrac{1}{2}\eta_{ecd(a}Dg_{b)}{}^e\wedge\Psi^c\wedge\omega^d
+\tfrac{1}{2} \eta_{rcd(b} \tilde R^r{}_{a)} \wedge \omega^c \wedge \omega^d
- \tfrac{1}{2} g_{ab} L_G. \label{tre}
\end{align}
 In the second equation we used, for $g^{ab}$ fixed, $\delta \tilde \omega_{ab}= \delta \omega_{[ab]}$.
 Actually these variational derivatives are not independent as the invariance under change of frame implies the identity
 (it is the Finslerian analog of the Einstein-Cartan formula \cite[Eq.\ (3)]{trautman71a}, see Appendix C).
\begin{equation} \label{toz}
\frac{\delta {L}_G}{\delta \omega^a} \wedge  \omega^b   =2 \frac{\delta {L}_G}{\delta g_{rb}}g_{ra}+D  \frac{\delta {L}_G}{\delta \omega^a{}_{ b}}.
\end{equation}
As a consequence, it is sufficient to consider the dynamical equations for the connection and vierbeins.

Further, it can be observed that under a variation of section
\[
\delta S_G=\int_M \delta s^*(L_G) ,
\]
but $L_G=\tilde \tau_d \wedge \omega^d$ thus, as in vacuum $\tilde \tau_d=0$, the action does not really depend on the section when the gravitational fields are on-shell.
It remains to establish the consequences of the vacuum equations which are obtained, by the arbitrariness of the section and  by the result of Appendix A, with the variational derivatives (\ref{one})-(\ref{tre}) set to zero.
Let
\begin{equation}
\tilde \omega^a{}_b=\tilde H^a{}_{bc} \omega^c+\tilde V^a{}_{bc} { \tilde{\omega}}^c,
\label{eq:omega-decomposition}
\end{equation}
where ${\tilde{\omega}}^c=\tilde D y^c$,
be the expansion for the connection coefficients.\footnote{This object should have been denoted $\bar{\tilde{\omega}}^c$, we hope that the simplification of notation does not cause confusion.}

Let $[e_a,e_b]=c_{ab}^c(x) e_c$
be the commutation relations for the chosen basis, so that $\dd e^c=-\tfrac{1}{2} c^c_{ab} e^a \wedge e^b$.
Then
\begin{equation} \label{eq:expand-Psi}
\Psi^a=\tilde D \omega^a=(\tilde H^a{}_{bc}+\tfrac{1}{2} c^a_{bc}) \omega^c \wedge \omega^b+\tilde V^a{}_{bc} {\tilde{\omega}}^c\wedge \omega^b.
\end{equation}

Let us first consider the consequences of the vacuum equation for the connection $\frac{\delta L_G}{\delta \omega^a{}_r} =0$, that is $\eta_{abcd}   \Psi^c  \wedge \omega^d=0$. It can be rewritten
\begin{equation} \label{bjs}
\Psi^a\wedge \omega^b-\Psi^b\wedge \omega^a=0
\end{equation}
 which implies\footnote{Because,  the algebraic equation $\delta^b{}_{[r} B^a{}_{pq]}-\delta^a{}_{[r}B^b{}_{pq]} = 0$, where $B^a_{bc}=-B^a_{cb}$, traced on the indices $b$ and $r$, and then traced again, easily leads to $B^a{}_{pq}=0$ (in spacetime dimension 4 and higher). Here $ B^a_{bc}:=H^a{}_{[bc]}+\tfrac{1}{2} c^a_{[bc]}$. }
 $\tildeHor( \Psi^a)=0$
 and hence $\tilde H^a{}_{[bc]}+\tfrac{1}{2} c^a_{[bc]}=0$.

But metricity of $\tilde D$  implies, in particular, $\tildeHor
 (\tilde  D g_{ab} )=0$, that is (by $\omega_c:=e_c^\mu \frac{\p}{\p x^\mu}- N^r_c(x,y) \frac{\p}
 {\p y^c}$ we denote the horizontal lift of $e_c$, by $\tilde \omega_c:=e_c^\mu \frac{\p}{\p x^\mu}- \tilde N^r_c(x,y) \frac{\p}
 {\p y^c}$ the tilde-horizontal lift of $e_c$, so that $\omega^b(\tilde \omega_c)=\omega^b(\omega_c)=\delta^b_c$, $\tilde{\omega}^b(\tilde \omega_c)=
 \tilde D_{\tilde{\omega}_c} y^b=0$)
\[
0=\tilde{\omega}_c (g_{ab})-\tilde H_{abc}-\tilde H_{bac}.
\]
Combining the previous equation with $\tilde H^a{}_{[bc]}+\tfrac{1}{2} c^a{}_{[bc]}=0$ we arrive at
\begin{equation} \label{jvt}
\tilde H^a{}_{bc}=\frac{1}{2}g^{ar}\{\tilde \omega_b(g_{ra})+\tilde \omega_a(g_{rb})-\tilde \omega_r(g_{ab}) -(c_{acb}+c_{bac}-c_{cba})\}.
\end{equation}
Let us now consider the vertical information in Eq.\ (\ref{bjs}).
The contraction $i_Y$ of Eq.\ (\ref{bjs}) where $Y$ is vertical
gives, setting $V^a_b:= \tilde V^a_{bc} Y^c$,
\[
\delta^b_{[r} V^a_{d]}-\delta^a_{[r} V^b_{d]}=0.
\]
Taking the trace we get $V^a_b=0$
so, by the arbitrariness of $Y$, $\tilde V^a_{bc}=0$, that is, $\Psi^a=0$. 
The equation $\tilde V^a{}_{bc}=0$  implies in turn that $\tilde D$ is strongly regular and that $\{\omega^c, \tilde \omega^c\}$ is the dual basis to $\{\tilde \omega_c, \frac{\p}{\p y^c}\}$, see  \cite{minguzzi21b} .
But $\tilde D$ is metric, so a calculation of the vertical part of $\tilde D g_{ab}=0$ gives $\frac{\p g_{ab}}{\p y^c}=0$, namely the metric does not depend on the vertical variables. Finally, the expression (\ref{jvt}) proves that in vacuum the horizontal coefficients of $\tilde \nabla$ do not depend on the vertical variables and are actually those of the Levi-Civita connection of $g_{ab}$.

In conclusion, the vacuum dynamical equation for the connection is equivalent to\footnote{This is the  Finslerian analog of the connection equation \cite[(38), see also (33)]{minguzzi20b} for our vacuum purely metric-affine theory. In that pseudo-Riemannian theory if metricity is imposed a priori, we are back to Einstein-Cartan theory and the connection vacuum equation states that the torsion vanishes. In the  Finslerian theory the equation has consequences also on the metric, not just on the connection.} $\Psi^a=0$  which is equivalent to the imposition of the following three conditions: (a) the independence of $g$ on the vertical variable, (b) the horizontal coefficients of $\tilde \nabla$ are those of the  Levi-Civita connection of $g$, (c) the vertical coefficients of $\tilde \nabla$ vanish  (we proved one direction, the other being clear).

As for the other vacuum dynamical equations, by Eq.\ (\ref{toz}) we need only to consider the vacuum dynamical equation for $\omega^a$, and this is $\tilde \tau_d=0$, which by the found form of the connection and Eq.\ (\ref{hhv}) is the statement that the Einstein tensor (and hence the Ricci tensor) of the Levi-Civita connection vanishes.


The original connection $\nabla$ can only be determined up to a gauge because we implemented the amplified symmetry. The same Lagrangian with all the `tilde' dropped would have established that the connection in vacuum is indeed Levi-Civita. It is indeed possible to impose that the connection is metric since the beginning. That gives a viable and physically equivalent approach, the only drawback being that the variation has to be constrained a priori since only metric connections should be considered (as in  the Einstein-Cartan theory). Therefore, it is not an approach in which metric and connection are completely unconstrained (purely metric-affine).

In any case, the conclusion is that in vacuum the theory is equivalent to general relativity, so according to the present theory in vacuum there is no torsion nor anisotropy. This seems a peculiar feature of this theory as most other   proposals for a Finslerian dynamics leave room for possible vacuum non-pseudo-Riemannian solutions \cite{hohmann19,pfeifer12,javaloyes21}. Still it is a natural one as our theory mimicks Einstein-Cartan,  in which, analogously, there is no torsion in vacuum. Also it fully agrees with our current experimental
evidence that vacuum is described by a pseudo-Riemannian (Lorentzian)
theory at the classical level.


\section{Conclusions}

We explored an anisotropic theory in which an Einstein-Cartan-like Lagrangian
on the slit tangent bundle $E$ is pulled back to $M$ through a section $s: M\to E$. The dynamics was obtained by imposing stationarity for every possible section $s$, which results in equations on $E$ independent of the section. This fact is interpreted as independence of the equations from the observer. The dynamical equations imply that in vacuum the theory is in fact coincident with general relativity and hence that there is no torsion nor anisotropy. The theory might also implement the amplified symmetry of our previous work. This modification, motivated by the purpose of interpreting the physical field as the non-linear connection, can also be omitted in which case the theory is framed in terms of a metric connection, this time living on $E$ rather than $M$, much as in the original Einstein-Cartan theory.

\section*{\large Appendix A: Proof of \ $\forall s, \, s^*\omega=0 \, \Rightarrow \, \omega=0$ }
\label{subsec:appendix-sstar}

In this appendix we provide a detailed argument on why $s^*\omega=0$ for every section $s: M\to E$ implies $\omega=0$. Here $\omega$ is a $k$-form on $E$, $k\le n$, $n$ dimension of $M$.

Let $E=TM\backslash 0$ and let $\pi:E\to M$ be the projection on the base.
The coordinates on $E$ are $(x^a,y^a)$, those on $TE$ are $(x^a, y^a, \dot x^a, \dot y^a)$. Elements in $TE$ are denoted with capital letters, e.g.\ $V$, those in $TM$ with lowercase letters, e.g.\ $v$.

Suppose $s^*\omega=0$ for every $s$. In order to prove that $\omega=0$ we just need to prove that $\omega$ vanishes over every $k$-vector $V_1\wedge V_2\wedge\cdots \wedge V_k$ with $V_i\in TE$.

We can assume that $v_i=\pi_*(V_i)$ are linearly independent and hence that
\[
\pi_*(V_1\wedge V_2\wedge\cdots \wedge V_k)=v_1\wedge v_2\wedge\cdots \wedge v_k \ne 0
\]
indeed, if not, replace $V_i$ with $V_i'(\epsilon)=V_i+\epsilon W_i$ for sufficiently small $\epsilon$ where $w_i=\pi_*(W_i)$ are linearly independent (if $t<k$ is the dimension of the space spanned by $\{v_i\}$ pick $t$ vectors of $\{w_i\}$ as a basis of this space and $k-t$ vectors as a basis for a trasverse subspace). Then $\{V_i'\}$ would have linearly independent projections, and if we can prove for these type of $k$-vectors that
\[
\omega(V'_1\wedge V'_2\wedge\cdots \wedge V'_k)=0,
\]
then taking the limit $\epsilon \to 0$ we also get
\[
\omega(V_1\wedge V_2\wedge\cdots \wedge V_k)=0.
\]

Let $P\in E$, we want to prove that $\omega(P)=0$. Let $p=
\pi(P)$. We need only to try the form on $V_1\wedge V_2\wedge\cdots \wedge V_k$ with projection $v_1\wedge v_2\wedge\cdots \wedge v_k\ne 0$, that is, we can assume that $v_i$ are linearly independent.

Introduce coordinates at the point $p\in M$ of interest such that $x^a(p)=0$. Consider the section $s: M\to E$ having local components
\[
s^a(x)=P^a+B^a{}_b x^b
\]
where $P^a=y^a(P)$ and $B$ is a constant $n\times n$ matrix. Note that
\[
V_i=v_i^a\frac{\p}{\p x^a}+q_i^b \frac{\p}{\p y^b}
\]
for some constants $q_i^b$.
Choose $B$ so that it maps $v_i^b\in \mathbb{R}^n$ to $q_i^b\in \mathbb{R}^n$, i.e.\ $q_i^b=B^b{}_a v_i^a$ (here linear independence of $\{v_i\}$ is used), then
\[
s_*(v_i)= v_i^a \frac{\p}{\p x^a}+ \frac{\p s^a}{\p x^c} v_i^c \frac{\p}{\p y^a}=V_i \in TE ,
\]
and hence
\[
\omega(V_1\wedge V_2\wedge\cdots \wedge V_k)=s^*\omega(v_1\wedge v_2\wedge\cdots \wedge v_k)=0.
\]

\section*{\large Appendix B: Proof of the expressions (\ref{one})-(\ref{tre})}
\label{subsec:appendix-identity}

The dynamical equations (\ref{one}) and (\ref{two}), that are the most important, can be proved directly.

Let us consider the variation just with respect to $\omega^a$.  We have
\[
\delta L_G= \eta_{abcd} \tilde R^{ab} \wedge \omega^c \wedge \delta \omega^d
\]
from which Eq.\ (\ref{one}) follows.

Let us consider the variation just with respect to $\omega^a{}_b$. From Eq.\ (\ref{byq})  $\delta \tilde \omega^a{}_b= \frac{1}{2} \delta \omega^a{}_b-\frac{1}{2}  g_{bs} \delta \omega^s{}_r g^{ra} $, that is $\delta \tilde \omega_{ab}= \delta \omega_{[ab]}$. We denote $g_{br}\frac{\delta }{\delta \omega^a{}_r}$ with  $\frac{\delta }{\delta \omega^{ab}}$. Now, the symmetric part  $\frac{\delta L_G }{\delta \omega^{(ab)}}$ vanishes due to the amplified symmetry (in the  version in which the tildes are dropped it does not make sense to consider this variation as $g_{ar}\delta \omega^r{}_{b}$ would be antisymmetric by the a priori compatibility of $D$ with the metric). Thus we have the identity  $\frac{\delta  L_G}{\delta \omega^{ab}}=\frac{\delta L_G}{\delta \omega^{[ab]}}=\frac{\delta L_G}{\delta \tilde \omega^{ab}}$.

But we have, using $\tilde D \omega^a=\Psi^a$,
\[
\delta L_G=\tfrac{1}{2} \eta_{abcd} \tilde D \delta \tilde \omega^{ab} \wedge \omega^c \wedge \omega^d=\dd [\tfrac{1}{2} \eta_{abcd}  \delta \tilde \omega^{ab} \wedge \omega^c \wedge \omega^d]- \eta_{abcd} \tilde \omega^{ab} \wedge \Psi^c \wedge \omega^d\wedge  \delta \tilde \omega^{ab},
\]
which proves Eq.\ (\ref{two}).

The last equation (\ref{tre}) is not really necessary for the dynamics as it follows from the other two, see Eq.\ (\ref{toz}). Anyway, we calculated its expression as follows, which also provides another proof of (\ref{one}) and (\ref{two}).

We start from the expression
$L_G=\tilde \tau_d \wedge \omega^d$. Then the variational derivatives follow from the identity
\begin{eqnarray}
&&
	\delta(\omega^c\wedge\tilde{\tau}_c)=
	d\left(\eta^{a}{}_{bcd} \omega^{c} \wedge \left(\frac{1}{2}\omega^{d}\wedge \delta\omega^{b}{}_{a} -
	\frac{1}{4}\delta g_{ae}\omega^{b} \wedge\mathcal{G}^{de}\right)\right)-\nonumber\\
&&
	-2 \tilde{\tau}_{a} \wedge \delta\omega^{a}
	+  \eta^{a}{}_{bcd} \Psi^{b}\wedge \omega^{d}\wedge \delta\omega^{c}{}_{a}\nonumber\\
&&
	-\frac{1}{2} \left(g^{ab} \tilde{\tau}_{c} \wedge \omega^{c} +
	\eta^{a}{}_{cde} \left(
		 \Psi^{c} \wedge \omega^{d} \wedge \mathcal{G}^{be}
		+\omega^{c} \wedge \omega^{d} \wedge\tilde{R}^{eb}
	\right)
\right)\delta g_{ab},
\label{eq:action-integral-variation}
\end{eqnarray}

To prove \eqref{eq:action-integral-variation} we will work out its left and right hand sides
and show that the obtained expressions agree.
Let us start by computing the left hand side: we expand out the expression $\omega^{c} \wedge \tilde{\tau}_{c}$
using the definition of $\tilde{\tau}_{c}$ given by \eqref{eq:tilde-tau} and the definition of $\tilde{R}_{ab}$ given by
\eqref{eq:tilde-R}. The result is, setting $\mathcal{G}_{ab}:=D g_{ab}$,
\begin{equation}
\omega^{c} \wedge \tilde{\tau}_{c} = \frac{1}{2} \eta_{acbd} \omega^{c} \wedge \omega^{b} \wedge R^{da} -
\frac{1}{8} \eta_{cdab} \omega^{c} \wedge \omega^{a} \wedge \mathcal{G}_{e}{}^{d} \wedge \mathcal{G}^{eb}.
\end{equation}
Next, we compute $\delta(\omega^{c} \wedge \tilde{\tau}_{c})$ (the left hand side
of \eqref{eq:action-integral-variation}) obtaining
\begin{eqnarray}
&&
	\delta(\omega^{c} \wedge \tilde{\tau}_{c}) = \nonumber\\
&&
	\tfrac{1}{2} \Bigl(\eta_{acbd} \bigl(\delta \omega^{c} \wedge \omega^{b} \wedge {R}^{da} +
	\omega^{c} \wedge \delta \omega^{b} \wedge {R}^{da} + \omega^{c} \wedge \omega^{b} \
	\wedge \delta {R}^{da}\bigr) + \nonumber\\
&&
	\omega^{c} \wedge \omega^{b} \wedge {R}^{da}\delta \eta_{acbd}\Bigr) +
	\tfrac{1}{8} \Bigl(-\omega^{c} \wedge \omega^{a} \wedge \mathcal{G}_{e}{}^{d} \wedge \mathcal{G}^{eb}\delta\eta_{cdab} - \nonumber\\
&&
	\eta_{cdab} \bigl(\delta\omega^{c}\wedge \omega^{a} \wedge \mathcal{G}_{e}{}^{d} \wedge \mathcal{G}^{eb} +
	\omega^{c} \wedge \delta\omega^{a}\wedge \mathcal{G}_{e}{}^{d} \wedge \mathcal{G}^{eb} +\nonumber\\
&&
	\omega^{c} \wedge \omega^{a} \wedge \mathcal{G}_{e}{}^{d} \wedge \delta\mathcal{G}^{eb} +
	\omega^{c} \wedge \omega^{a} \wedge \delta\mathcal{G}_{e}{}^{d}\wedge \
\mathcal{G}^{eb}\bigr)\Bigr)
\label{eq:delta-omega-tau}
\end{eqnarray}

The procedure is now a straightforward but tedius computation that involves
replacing the variations of $g^{ab}$, $\eta_{abcd}$, $\mathcal{G}_{ab}$ and $R^{a}{}_{b}$ using the following
relations
\begin{eqnarray}
 &&
  \delta g^{ab} = - g^{ac} g^{bd} \delta g_{cd}\;,\quad
 \delta \eta_{abcd} = \frac{1}{2} \eta_{abcd} g^{eh} \delta g_{eh}\;,\quad
 \delta R^{a}{}_{b} = D\delta \omega^{a}{}_{b}\;,\quad\label{eq:var1}\\
 &&
\delta\mathcal{G}_{ab} = D\delta g_{ab} -  g_{bc} \delta \omega^{c}{}_{a} -  g_{ac} \delta \omega^{c}{}_{b}.
\label{eq:var2}
\end{eqnarray}

Now, one can replace the exterior derivative that appears on the
right hand side of \eqref{eq:action-integral-variation} by the exterior covariant derivative $D$ getting
\begin{eqnarray*}
&&
	D\left(\eta^{a}{}_{bcd} \omega^{c} \wedge \left(\frac{1}{2}\omega^{d}\wedge \delta\omega^{b}{}_{a}
	-\frac{1}{4}\delta g_{ae}\omega^{b} \wedge\mathcal{G}^{de}\right)\right)=\\
&&
	-\frac{1}{2} \eta_{abcd} \delta\omega^{a}{}_{e}\wedge \omega^{b} \wedge \omega^{c} \wedge \mathcal{G}^{de} +\nonumber\\
&&
	\tfrac{1}{4}\eta^{a}{}_{bcd} \Bigl(2 D\delta \omega^{b}{}_{a} \wedge \omega^{c} \
	\wedge \omega^{d} + 4 D\omega^{b}\wedge \delta\omega^{c}{}_{a} \
	\wedge \omega^{d} -  \omega^{b} \wedge \omega^{c} \wedge \
	\mathcal{G}^{de} \wedge D\delta g_{ae}\Bigr) + \nonumber\\
&&
	\tfrac{1}{4}\delta g_{ab}
	\Bigl(\eta^{a}{}_{cde} \bigl(2 D\omega^{c} \wedge \omega^{d} \wedge \mathcal{G}^{be} + \omega^{c} \wedge \omega^{d} \
	\wedge D\mathcal{G}^{be}\bigr) -  \eta_{cdef} \omega^{c} \wedge \omega^{d} \wedge \mathcal{G}^{ae} \wedge \mathcal{G}^{bf}\Bigr)-\nonumber\\
&&
	\tfrac{1}{4}g^{ab} \bigl(2 \delta\omega^{c}{}_{a} \wedge \omega^{d} \wedge \omega^{e} \wedge D\eta_{bcde}
+ \delta g_{ac} \omega^{d} \wedge \omega^{e} \wedge \mathcal{G}^{cf} \wedge D\eta_{bdef}\bigr),
\end{eqnarray*}
and compute the latter using the
following \emph{structure equations}
\begin{eqnarray}
&&
	D\omega^{a} = T^{a},\\
&&
	D\eta_{abcd} = \frac{1}{2} \mathcal{G}^{e}{}_{e} \eta_{abcd},\\
&&
	D\mathcal{G}^{cr} = -2 R^{(cr)}.
\end{eqnarray}
After a long computation one obtains an expression for the right hand side
of \eqref{eq:action-integral-variation}
that agrees with the value of $\delta(\omega^{c} \wedge \tilde{\tau}_{c})$ obtained from
\eqref{eq:delta-omega-tau} after using \eqref{eq:var1}-\eqref{eq:var2}.

\section*{\large Appendix C: Proof of identity \eqref{toz}}
\label{subsec:appendix-toz}
Let $A^a{}_b$ be the \emph{transition functions} obtained from
suitable
\emph{local trivializations} of the bundle $\pi^*(E)$ and define
$\bar{A}^{a}{}_b$
by the relation
$A^a{}_b\bar{A}^b{}_c=\bar{A}^a{}_bA^b{}_c=\delta^a{}_c$.
The \emph{cocycle relation} of $\omega^a$ is then given by
\begin{equation}
 \omega'^a=\bar{A}^a{}_b\omega^b.
\label{eq:transition-omega}
\end{equation}
This induces the following cocycle relations for $\omega^a{}_b$ and
$g_{ab}$
\begin{equation}
 \omega'^r{}_b=\bar{A}^r{}_a\omega^a{}_sA^s{}_b-
A^c{}_b\dd(\bar{A}^r{}_c)\;,\quad
 g'_{ab}=g_{cd} A^c{}_a A^d{}_b.
\label{eq:transition-connection}
\end{equation}

Let us
assume that we take a family of transition functions depending on a
parameter $t$,
$\{(A^t)^a{}_b, (\bar{A}^t)^a{}_b\}$, and define
$\dot{A}^{a}{}_{b}:=\tfrac{d}{dt}(A^t)^{a}{}_b\rvert_{t=0}$,
 $\dot{\bar{A}}^{a}{}_{b}:=\tfrac{d}{dt}(\bar{A}^t)^{a}{}_b\rvert_{t=0}
$.
Then if we take the derivative with respect to $t$
of \eqref{eq:transition-omega}-\eqref{eq:transition-connection}
particularized for the family of transition functions just defined, we
get
\begin{equation}
 \delta \omega^a = \dot{\bar{A}}^a{}_b \omega^b\;,\quad
 \delta \omega^a{}_b =\omega^a{}_c\dot{A}^c{}_b-
\dd(\dot{\bar{A}}^a{}_b)+\dot{A}^a{}_c\omega^c{}_b\;,\quad
 \delta g_{ab} = g_{cb}\dot{A}^c{}_a+g_{ad}\dot{A}^d{}_b,
\label{eq:variations}
\end{equation}
where, as usual, we define the variations of the configuration
variables $\omega^a$, $\omega^a{}_b$, $g_{ab}$ adapting
\eqref{eq:delta-variation} to the present situation. Next we assume
that we have an action $S=\int_Ms^*\mathcal{L}$, where
$\mathcal{L}$ is a function of $\omega^a$, $\omega^a{}_b$, $g_{ab}$ and
it does not change under \eqref{eq:transition-omega}. The
general variation $\delta S$ is given by
\begin{eqnarray}
&&
	\delta S=\int_M s^*\left(\frac{\delta{\mathcal L}}{\delta
\omega^a}\wedge \delta\omega^a\right)+
	\int_M s^*\left(\frac{\delta\mathcal L}{\delta \omega^a{}_b}
\wedge \delta \omega^a{}_b\right)+
	\int_M s^*\left(\frac{\delta{\mathcal L}}{\delta
g_{ab}}  \delta g_{ab}\right)\nonumber\\
&&
	+\int_M s^*(\dd \Xi).
\end{eqnarray}
If we particularize the above for the variation induced by the
transformation \eqref{eq:transition-connection}, then, on
one hand $\delta S=0$, and on the other, we may replace $\delta
\omega^a$, $\delta \omega^a{}_b$, $\delta g_{ab}$ by the values
given by \eqref{eq:variations}. Using integration by parts in the
resulting expression, the identity
\begin{equation}
	\dd\left(\frac{\delta\mathcal L}{\delta
\omega^a{}_b}\dot{\bar{A}}^a{}_b \right) =
	D(\dot{\bar{A}}^a{}_b)\wedge\frac{\delta\mathcal L}{\delta
\omega^a{}_b}+
	\dot{\bar{A}}^a{}_b D\left(\frac{\delta\mathcal L}{\delta
\omega^a{}_b}\right),
\end{equation}
and the the Stokes theorem on the boundary
terms we deduce
\begin{equation}
	0=\int_M\dot{\bar{A}}^r{}_s s^*\left(\frac{\delta
\mathcal{L}}{\delta\omega^r}\wedge\omega^s
	-2g_{rb}\frac{\delta \mathcal{L}}{\delta g_{sb}}
	-D\left(\frac{\delta\mathcal L}{\delta
\omega^r{}_s}\right)\right).
\end{equation}
Given that the section $s$ is arbitrary, the term in brackets must
vanish, leading straight to \eqref{toz}.


\section*{Acknowledgments}
A.G.P.\ is supported by projects PY20-01391 and UCO-1380930  from
the Regional Government of Andalusia (Spain) and ERDEF (UE). E.M.\ is partially supported by GNFM of INDAM.\\


\end{document}